\documentclass[aps, twocolumn, tightenlines, ece, amssymb, amsfonts, amsmath, groupedaddress]{revtex4}
\usepackage{epsf}
\usepackage{graphics}
\usepackage{graphicx}
\usepackage{epstopdf}

\begin{document}

\title{Formation and magnetic properties of spark plasma sintered Mn$_{3 - \delta}$Ga ($\delta$ = 0, 1) alloys}

\author{Sonam Perween$^{a,b}$, A. Rathi$^{a,b}$, R. P. Singh$^{c}$, A. Bhattacharya$^{d}$, Parul Rani Raghuvanshi$^{d}$,
P. V. Prakash Madduri$^{e}$, P. K. Rout$^{a}$, B. Sivaiah$^{a,b}$,
Ajay Dhar$^{a,b}$, R. P. Pant$^{a,b}$, B. Gahtori$^{a,b,\dag}$ and
G. A. Basheed$^{a,b,\dag}$}

\affiliation{$^{a}${CSIR-National Physical Laboratory (NPL), Dr. K.
S. Krishnan Marg, New Delhi - 110012, India}\\
$^{b}${Academy of Scientific and Innovative  Research (AcSIR), NPL
Campus, New Delhi - 110012, India}\\
$^{c}${Department of Physics, Indian Institute of Science Education
and Research (IISER) Bhopal, Bhopal - 462023, India}\\
$^{d}${Department of Metallurgical Engineering and Materials Science, IIT Bombay, Mumbai-400076, India}\\
$^{e}${School of Physics, University of Hyderabad, Central
University P. O., Hyderabad 500046, Telangana, India}}

\email[$\dag$]{basheedga@nplindia.org, bhasker@nplindia.org}

\begin{abstract}

We present the synthesis of D0$_{22}$ Mn$_{3 - \delta}$Ga ($\delta$
= 0, 1) Heusler alloys by Spark Plasma Sintering method. The single
phase Mn$_3$Ga (T$_\mathrm{c}$ $\simeq$ 780 K) is synthesized, while
Mn$_2$Ga (T$_\mathrm{c}$ $\simeq$ 710 K) is found to coexist with a
near-stoichiometric room temperature paramagnetic
Mn$_9$Ga$_5$~($\approx$ 15 \%) phase due to its lower formation
energy, as confirmed from our density functional theory (DFT)
calculations. The alloys show hard magnetic behavior with large room
temperature spontaneous magnetization m$_s$(80 kOe) = 1.63 (0.83)
$\mu_\mathrm{B}$/f.u. and coercivity H$_\mathrm{c}$ = 4.28 (3.35)
kOe for Mn$_3$Ga (Mn$_2$Ga). The magnetic properties are further
investigated till T$_\mathrm{c}$ and the H$_\mathrm{c}$ (T) analysis
by Stoner-Wohlfarth model shows the nucleation mechanism for the
magnetization reversal. The experimental results are well supported
by DFT calculations, which reveal that the ground state of D0$_{22}$
Mn$_2$Ga is achieved by the removal of Mn-atoms from full Heusler
Mn$_3$Ga structure in accordance with half Heusler alloy picture.

\vspace{1.0em} \noindent{\it\textbf{Keywords}}: Mn-based alloys,
Heusler alloys, Coercivity mechanism, Density functional theory.

\end{abstract}
\maketitle

\section{INTRODUCTION}
\vspace{-1.0em}

The Mn-based Heusler alloys
\cite{PhysRevB.90.214420,PhysRevB.92.064417} have gained extensive
interest in magnetism research community because of their versatile
magnetic behavior and consequent use in multi-functional
applications, starting from spintronics to hard magnets. One such
family is the Mn$_{3 - \delta}$Ga alloys in the tetragonal D0$_{22}$
crystal structure
\cite{PhysRevB.92.064417,PhysRevB.77.054406,doi:10.1063/1.362115,doi:10.1002/pssb.201147122}
having a high Curie temperature (T$_\mathrm{c}$ $>$ 600 K), large
magnetic anisotropy K$_\mathrm{u}$ $\sim$ 10$^6$ erg/cc and high
spin polarization. Owing to these properties, the Mn-Ga family has
emerged as a favorable system for hard magnets (in bulk
\cite{PhysRevB.77.054406,doi:10.1063/1.362115}) and
spin-transfer-torque memory applications (in epitaxial thin films
\cite{doi:10.1002/pssb.201147122}). However, the synthesis of
D0$_{22}$ Mn$_{3 - \delta}$Ga (0 $\ge$ $\delta$ $\ge$ 1) alloys has
been a challenging task due to the existence of manifold stable
near-stoichiometric and structural phases \cite{Meissner1965340,
Lu1980469, MINAKUCHI2012332}. Till date, only arc melting technique
has been used to synthesize bulk D0$_{22}$ Mn$_{3 - \delta}$Ga
($\delta$ = 0, 1) alloys
\cite{PhysRevB.77.054406,doi:10.1063/1.362115,QMLu2015,doi:10.1063/1.4759351,PERWEEN2019278,doi:10.1063/1.4942557},
which show large discrepancies in the magnetic properties.
Meanwhile, the magnetism in hard-magnetic alloys is largely
controlled by the microstructure (grain size and homogeneity) of the
sample \cite{doi:10.1063/1.328996,KRONMULLER1988291}. For better
homogeneity, the arc-melting procedure needs to be repeated several
times and further annealing for a long time (i.e. 1-2 weeks) is
required \cite{PhysRevB.77.054406,doi:10.1063/1.362115}. Besides,
high evaporation rate of Mn makes it difficult for controlled and
reliable synthesis of these alloys. Moreover, the preparation of a
denser sputtering target requires hot-compression of arc-melted
sample, which also leads to a much higher coercivity
\cite{MIX201589}. To avoid aforementioned issues, a better
alternative route to synthesize these alloys is single-step Spark
Plasma Sintering (SPS) technique \cite{SPSbook2014,SPSbook2013}. In
addition to quick sintering and annealing temperatures, the
quasi-static compressive stress in SPS leads to maximum uniformity
and high mechanical strength \cite{Munir2006}, which are highly
advantageous for both hard magnets \cite{NdFeBSPS2007} and
sputtering targets \cite{SPSbook2013,SPStarget2019}.

The parent Mn$_3$Ga alloy is established as a full-Heusler alloy
(XX$^\prime$YZ) with a D0$_{22}$ structure, which is a tetragonal
variant of cubic L2$_1$ (\emph{Fm-3m}) phase due to strong lattice
distortion along c-axis \cite{GRAF20111}. This stabilizes
\emph{I4/mmm} crystal symmetry, where two unique Mn (X/X'- 4d and Y
- 2b) sites with an opposite arrangement of spins lead to a
ferrimagnetic structure. In this picture, D0$_{22}$ Mn$_{3 -
\delta}$Ga (0 $\ge$ $\delta$ $\ge$ 1) structure must be realized by
creating Mn vacancies at 4d-site such that Mn$_2$Ga ($\delta$ = 1)
is a half-Heusler alloy (XYZ). However, the \emph{ab initio}
calculations \cite{PhysRevB.77.054406} based on Heusler picture did
not account the experimental results on DO$_{22}$ Mn$_{3 -
\delta}$Ga  (0 $<$ $\delta$ $\le$ 1), and suggested the vacancy
formation at both Mn sites. Moreover, the neutron diffraction
studies on  Mn$_{3 - \delta}$Ga thin film \cite{PhysRevB.85.014416}
and ribbons \cite{ZHAO20176} suggest a substitution model with
random site occupation of Mn (2b) and Ga (2a) atoms, which occurs
for the tetragonal variant of B2-type structure \cite{GRAF20111}.
Thus, the most intriguing issue is to identify the Mn-atoms (2b or
4d or both) that leave Mn$_3$Ga lattice to form a Mn$_2$Ga crystal
structure and hence, to identify the ground state for Mn$_2$Ga.

Here, we synthesized the D0$_{22}$ Mn$_{3 - \delta}$Ga ($\delta$ =
0, 1) alloys using SPS technique for the first time. The magnetic
properties of two alloys are investigated over wide temperature
range (300 K to 950 K). We further employ density functional theory
calculations to systematically study the formation and magnetic
ground state of Mn$_{3 - \delta}$Ga ($\delta$ = 0, 1) alloys, which
are found to be consistent with our experimental results.

\vspace{-1.0em}
\section{Experimental Details}
\vspace{-1.0em}

The Mn$_3$Ga and Mn$_2$Ga alloys are synthesized by Spark Plasma
Sintering (SPS Syntex, 725) method. The stoichiometric amount of
high purity Mn (99.95\%) and Ga (99.99\%) are intermixed in glove
box with an excess of 4 wt.\% Mn to compensate the evaporation loss.
The sintering is performed at 800 $^\circ$C under a fixed pressure
of 8.5 kN for Mn$_3$Ga with the heating and cooling rates of
100$^\circ$C/min. A similar sintering process for Mn$_2$Ga is
carried out at 800 $^\circ$C under 6 kN pressure. The room
temperature X-ray diffraction (XRD) patterns are recorded using
Rikagu Miniflex diffractometer and analyzed with Rietveld method
using FULLPROF program. The as-sintered Mn$_3$Ga shows mixed phase
with hexagonal and D0$_{22}$ tetragonal structure whereas D0$_{22}$
tetragonal Mn$_2$Ga with  $\approx$ 30 \% of Mn$_9$Ga$_5$ phase is
observed in as-sintered Mn$_2$Ga alloy. The differential scanning
calorimetry (DSC) measurements are performed on as-sintered alloys
to determine the annealing temperature for D0$_{22}$ phase
formation. Following this, the sintered pellets are vacuum sealed in
quartz tubes and annealed for 4 days at 725$^\circ$C for Mn$_3$Ga
and 400$^\circ$C for Mn$_2$Ga with the heating and cooling rates of
10$^\circ$C/min. The Energy-dispersive X-ray spectroscopy (EDS)
measurements give an average stoichiometry of Mn$_{2.90(5)}$Ga and
Mn$_{2.05(5)}$Ga for Mn$_3$Ga and Mn$_2$Ga samples, respectively.
The detailed magnetic measurements are performed by physical
properties measurement system (Quantum design) in the temperature
range of 300 K to 950 K.

\vspace{-1.0em}
\section{Calculational Details}
\vspace{-1.0em}

The DFT calculations are carried out using VASP
\cite{PhysRevB.49.14251}, a plane wave based electronic structure
code with projected augmented wave potential
\cite{PhysRevB.50.17953}. Perdew-Burke-Ernzerhof (PBE)
\cite{PhysRevLett.78.1396} exchange correlation functional within
generalized gradient approximation (GGA) are employed. An energy cut
off of 500 eV is used. The k-mesh is generated by Monkhorst-Pack
method and the convergence of the results are tested by varying the
mesh size. For the generation of the electronic density of
states~(DOS), denser k-grids are used. In all our calculations,
self-consistency is achieved with numerical settings that yield a
convergence for energy differences to $<10^{-3}$ eV/atom. The atomic
as well as geometrical optimization are performed via conjugate
gradient minimization \cite{PULAY1980393} and the forces on the
atoms are converged to less than 0.001 eV/$\mathrm{\AA}$. The
vibrational free energy of the compositions is calculated using
phonopy \cite{TOGO20151}.

\begin{figure}
\vspace{-1.0em}
\includegraphics[scale=0.40]{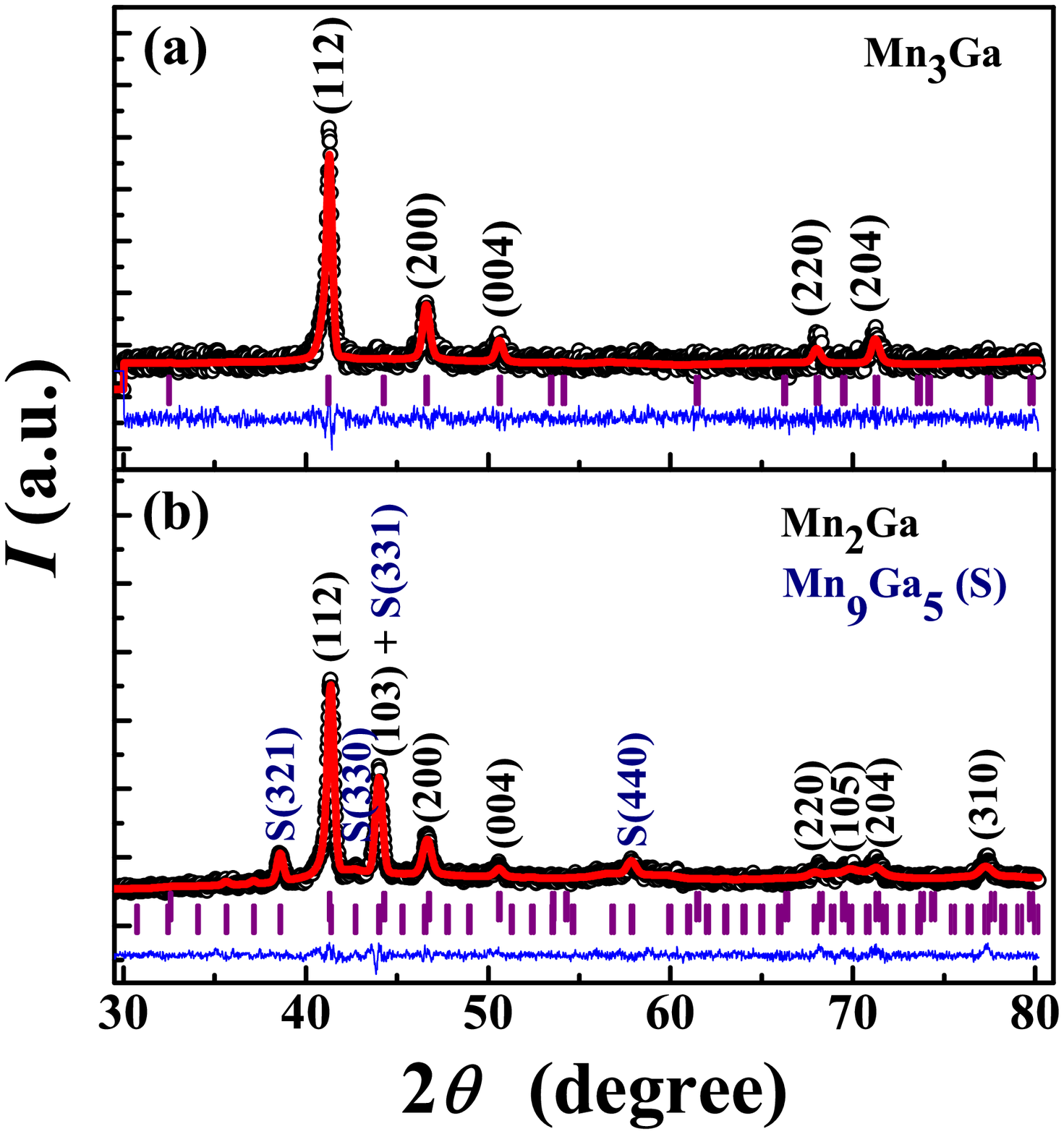}
\includegraphics[scale=0.37]{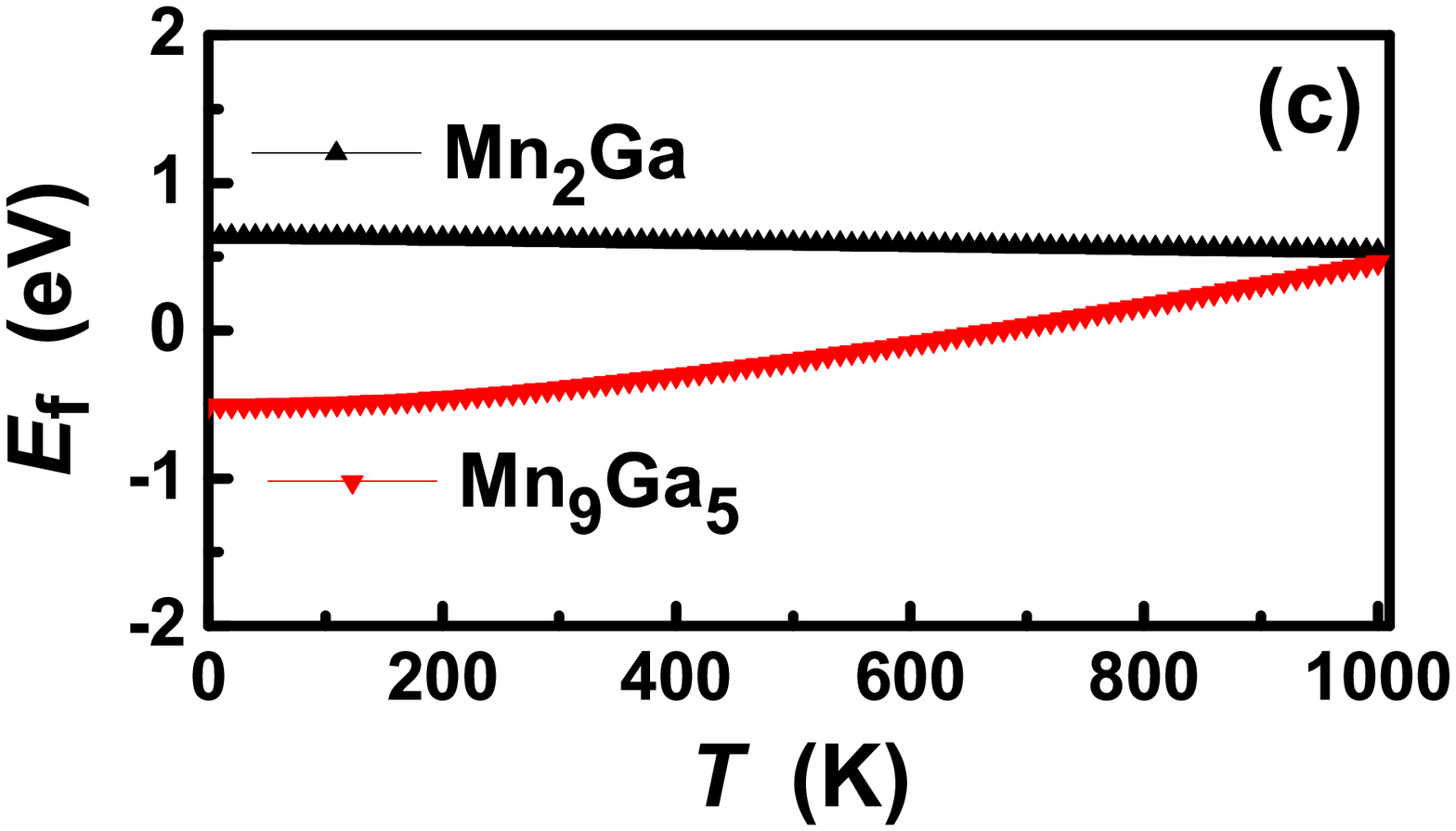}
\caption{Room temperature X-ray diffraction (XRD) patterns of (a)
Mn$_3$Ga and (b) Mn$_2$Ga, fitted with single-phase \emph{I4/mmm}
symmetry and two-phases i.e. main phase \emph{I4/mmm} and secondary
phase(S) \emph{P-43m} symmetries, respectively using Rietveld
method. (c) The thermal variation of the formation energy
$E_\mathrm{f}$ for near-stoichiometric Mn$_2$Ga and Mn$_9$Ga$_5$
phases.} \vspace{-1.0em}
\end{figure}

\vspace{-1.0em}
\section{Result and discussion}
\vspace{-1.0em}

\subsection{Structural Characterization}
\vspace{-1.0em}

The room-temperature XRD patterns of SPS and vacuum annealed
Mn$_3$Ga and Mn$_2$Ga alloys are shown in Fig. 1(a) and (b),
respectively. The Rietveld refinement confirms the single tetragonal
D0$_{22}$ phase (space group - \emph{I}4/mmm) for Mn$_3$Ga with
lattice parameters, a = 3.897(1) $\mathrm{\AA}$ and c = 7.213(1)
$\mathrm{\AA}$. On the other hand, the Mn$_2$Ga alloy is formed in
tetragonal phase [a = 3.890(1) $\mathrm{\AA}$, c = 7.205(1)
$\mathrm{\AA}$] with $\approx$ 15 \% volume fraction of
near-stoichiometric secondary Mn$_9$Ga$_5$ phase (P-43m symmetry).
In comparison to lattice parameters [a = 3.904(5) $\mathrm{\AA}$, c
= 7.091(8) $\mathrm{\AA}$] for arc-melted Mn$_3$Ga
\cite{PhysRevB.77.054406,doi:10.1063/1.4866844}, the \emph{a} for
SPS Mn$_3$Ga is quite similar whereas \emph{c} is notably larger.
Furthermore, previous arc-melted studies show an increase of
\emph{c} parameter for Mn$_2$Ga compared to Mn$_3$Ga
\cite{PhysRevB.77.054406,doi:10.1063/1.4866844}. In contrast, our
SPS samples reveal a slight decrease of 'c' parameter from Mn$_3$Ga
to Mn$_2$Ga, which is consistent with D0$_{22}$ Heusler alloy
picture with partial removal of 4d Mn atoms from Mn$_3$Ga unit cell
\cite{GRAF20111}.

To understand the formation of secondary Mn$_9$Ga$_5$ phase during
the synthesis of Mn$_2$Ga alloy, we performed DFT calculation of
formation energy \emph{E}$_f$ for two compositions. Figure 1(c)
shows the calculated \emph{E}$_f$ of Mn$_2$Ga and Mn$_9$Ga$_5$ with
respect to bulk phases by including the contributions stemming from
the vibrational free energy. The \emph{E}$_f$ for Mn$_9$Ga$_5$ comes
out to be smaller than that of Mn$_2$Ga at the sintering
temperature. This makes the exclusion of secondary Mn$_9$Ga$_5$
phase quite difficult during the synthesis of Mn$_2$Ga.
Nevertheless, Mn$_9$Ga$_5$ (T$_\mathrm{c}$ $<$ 165 K
\cite{TOZMAN2016147}) is non-magnetic in the temperature range
(300-950 K) under study and thus do not contribute to the magnetic
properties of Mn$_2$Ga.

\begin{figure}
\vspace{-1.5em}
\includegraphics[scale=0.35]{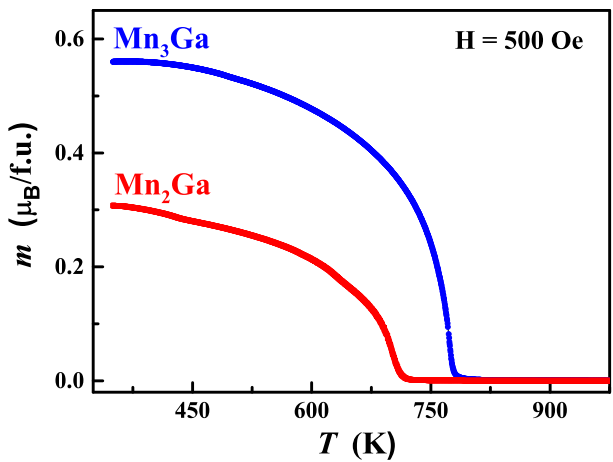}
\vspace{-2.5em} \caption{The temperature dependence of dc
magnetization for Mn$_3$Ga (blue) and Mn$_2$Ga (red) alloys in an
applied magnetic field of 500 Oe.} \vspace{-1.0em}
\end{figure}

\begin{figure}
\vspace{-2.0em}
\includegraphics[scale=0.35]{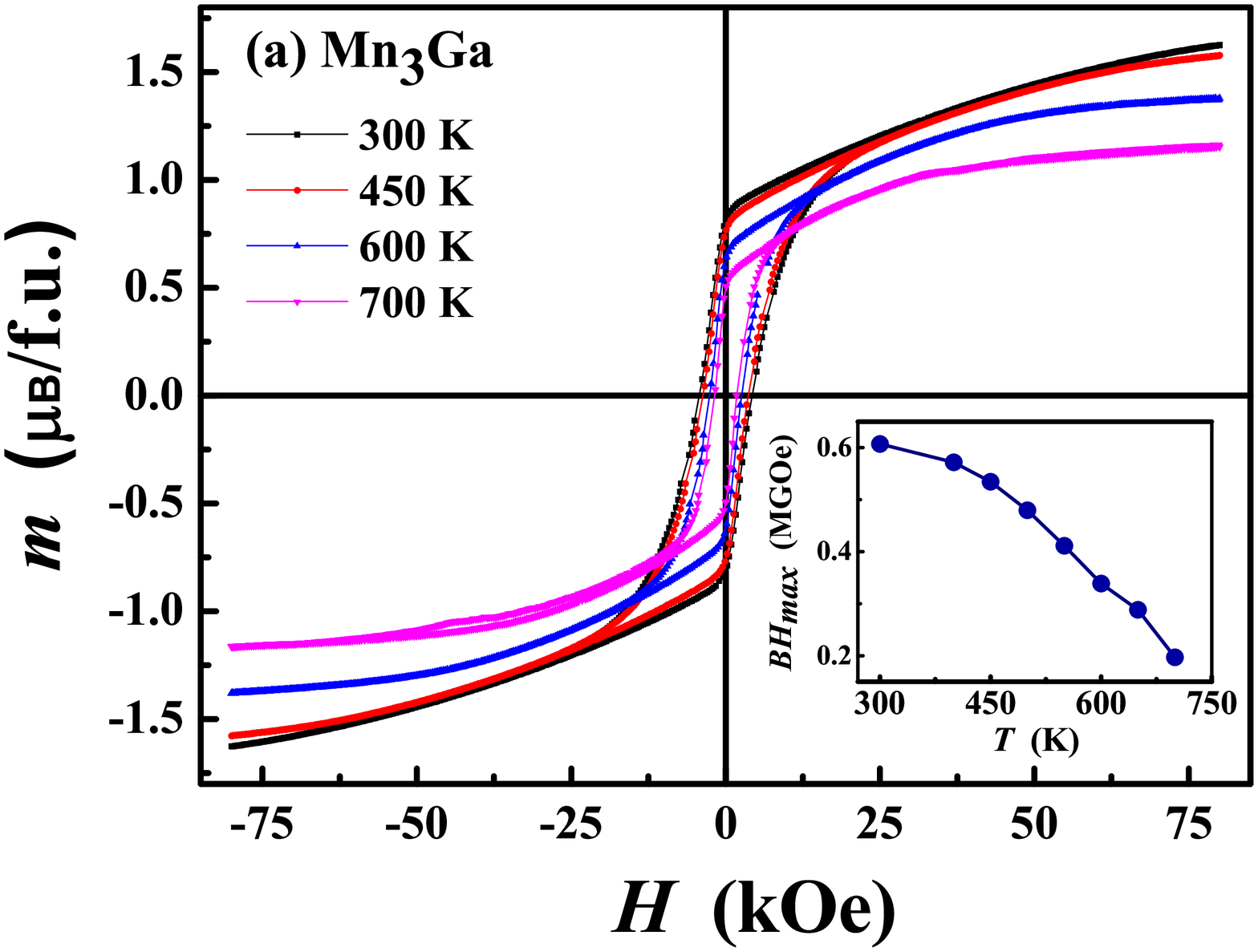}
\includegraphics[scale=0.35]{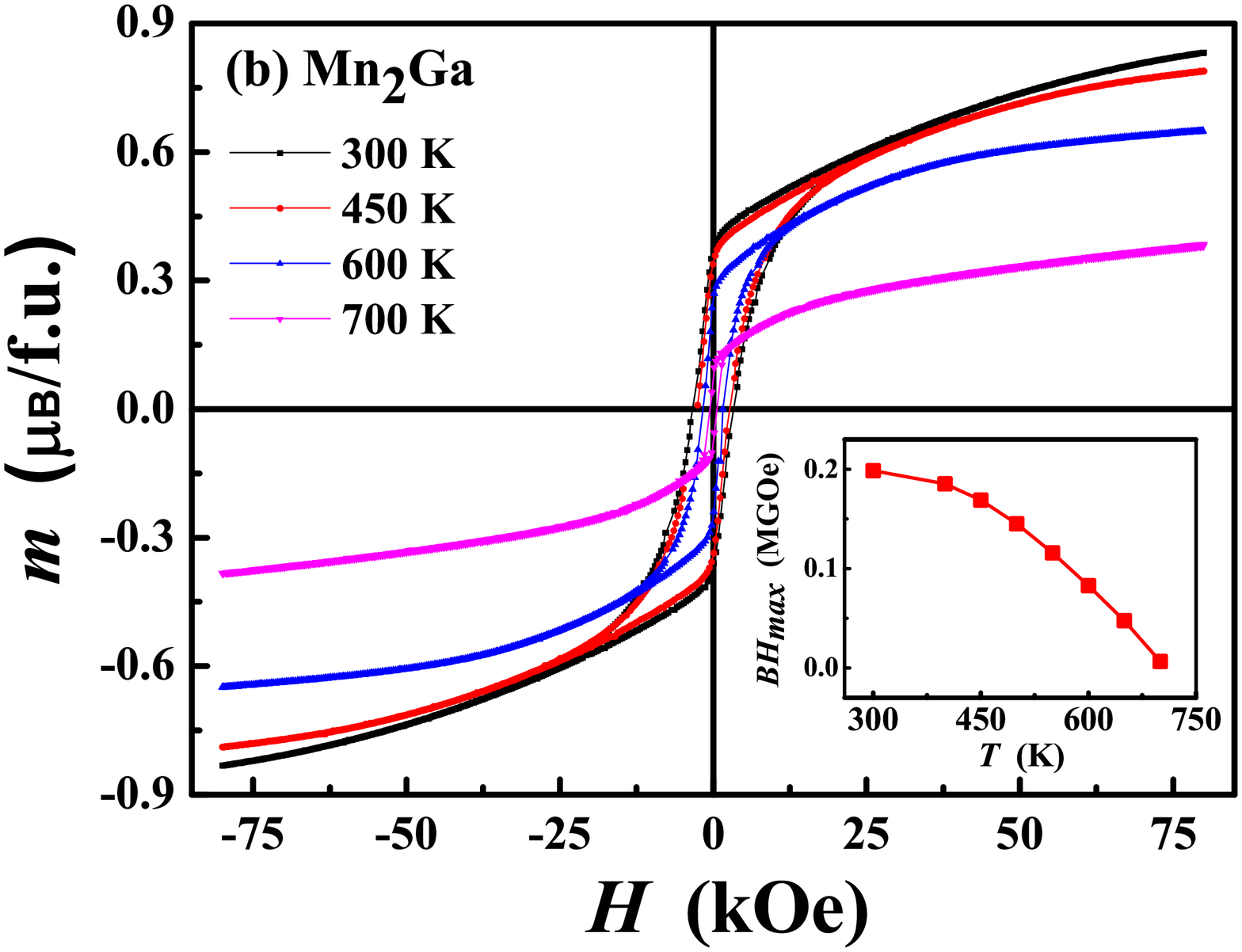}
\vspace{-2.0em} \caption{The field dependant magnetization curves of
(a) Mn$_2$Ga and (b) Mn$_3$Ga for selected temperatures. The inset
shows the thermal variation of maximum energy product
(BH$_\mathrm{max}$).} \vspace{-1.0em}
\end{figure}

\vspace{-1.0em}
\subsection{Magnetic Characterization}
\vspace{-1.0em}

Figure 2 shows the temperature dependence of dc magnetization,
$\emph{m}$(T) for Mn$_3$Ga and Mn$_2$Ga. The $\emph{m}$(T) show a
sharp increase below T$_\mathrm{c}$ $\simeq$ 780 K and $\simeq$ 710
K for Mn$_3$Ga and Mn$_2$Ga, respectively, which is consistent with
previous arc-melting study \cite{PhysRevB.77.054406}. Complementing
XRD study, the $\emph{m}$(T) results unambiguously confirm the
successful synthesis of D0$_{22}$ Mn$_{3 - \delta}$Ga alloys.

The field dependent magnetization, $\emph{m}$(H) loops for Mn$_3$Ga
and Mn$_2$Ga at selected temperatures are depicted in Fig. 3. The
$\emph{m}$(H) loops exhibit a broad hysteresis with non-saturating
behavior up to 80 kOe, which can be associated with a ferrimagnetic
structure and strong magnetic anisotropy present in these hard
magnetic alloys. The room temperature spontaneous magnetization
(m$_\mathrm{s}$) is observed to be 1.63 $\mu_\mathrm{B}$ at 80 kOe
(1.45 $\mu_\mathrm{B}$ at 50 kOe) for Mn$_3$Ga, which is
significantly larger than $\sim$ 1.1 $\mu_\mathrm{B}$ (at 50 kOe)
reported for arc melted alloy
\cite{PhysRevB.77.054406,doi:10.1063/1.2722206}. This may be
attributed to significantly larger 'c' parameter in SPS Mn$_3$Ga, as
Mn-Mn exchange coupling increases with interatomic distance
\cite{doi:10.1143/JPSJ.59.273}. In comparison, the room temperature
m$_\mathrm{s}$ for Mn$_2$Ga is 0.83 $\mu_\mathrm{B}$ at 80 kOe,
which is nearly half of the Mn$_3$Ga value. This is directly linked
to the structural evolution from Mn$_3$Ga to Mn$_2$Ga with the
formation of Mn-vacancies (either at 4d or 2b or both the sites).
Moreover, on increasing the temperature, the m$_\mathrm{s}$
decreases in a typical Bloch$'$s law manner
\cite{0295-5075-106-1-17001}, which confirms the long-range
ferrimagnetic ordering in these alloys. Here, we want to point out
that in contrast to present SPS study, the previous arc-melting
study \cite{PhysRevB.77.054406} shows a increase of moment from
Mn$_3$Ga to Mn$_2$Ga, which has been modeled by an \emph{ab initio}
study with removal of Mn-atoms from both the Mn-sites (one-third
from 4d-site and two-third from 2b-site). Such contrasting behavior
may stem from the difference in growth techniques. Unlike the
formation of alloys ``freely" in ``arc-melting" technique, the
``SPS" method involves large quasi-static compressive stress, which
can result in different atomic arrangement within the unit cell, in
particular for partially-filled structures like Mn$_2$Ga. This issue
is further analyzed using DFT calculations in section 4.3.

We now turn to the hard magnetic behavior in two alloys. The
$\emph{m}$(H) loops also shows a large room temperature coercivity
(H$_\mathrm{c}$) of 4.29 kOe (3.35 kOe) for Mn$_3$Ga (Mn$_2$Ga),
which monotonically decreases with temperature [See Fig. 4]. The
hardening of magnetic materials is generally dependent on the
magnetic anisotropy in the system, which acts against the coherent
rotation of the magnetic moments, according to Stoner-Wohlfarth
model \cite{Stoner599}. We have estimated the effective magnetic
anisotropy by fitting $\emph{m}$(H) isotherms to the law of {\it
``approach to saturation''} (ATS) \cite{KAUL2016539,RATHI2019585}
given as: $m(T, H) = \chi_{hf} \times H + m_{sat} [1 - b(T)/H^2]$.
Here, $\chi_{hf}$ represents high field susceptibility due to
paraprocesses and the coefficient $b(T)$ relates to the magnetic
anisotropy in the system. The nature of magnetic anisotropy can be
qualitatively determined from the remanence ratio
(m$_\mathrm{r}$/m$_\mathrm{s}$), which is found to be $\sim$ 0.49
(0.44) for Mn$_3$Ga (Mn$_2$Ga) alloy. The
m$_\mathrm{r}$/m$_\mathrm{s}$ values for two alloys are very close
to 0.5 as expected for uniaxial anisotropy. This indicates the
presence of uniaxial magneto-crystalline anisotropy (K$_\mathrm{u}$)
for two alloys. In that case, $b(T)$ is related to K$_\mathrm{u}$(T)
by the relation, $b(T)$ = 4 K$_\mathrm{u}^2$(T)/15 m$_{sat}^2$(T)
\cite{RATHI2019585}.

\begin{figure}
\vspace{-1.5em}
\includegraphics[scale=0.35]{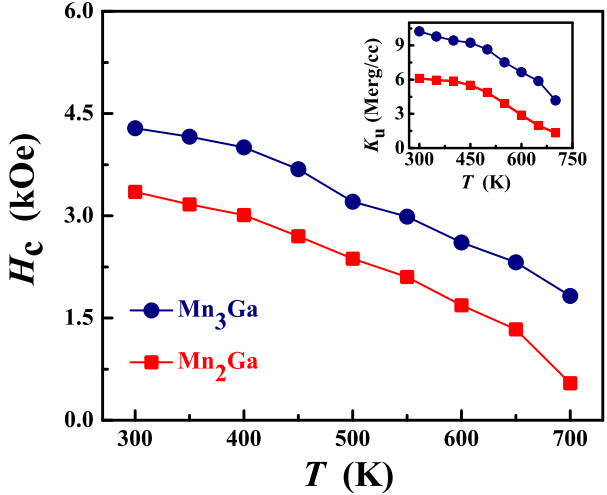}
\vspace{-2.0em} \caption{The temperature dependence of coercivity
($\emph{H}_c$) for Mn$_3$Ga (blue circles) and Mn$_2$Ga (red
squares). The inset shows the temperature dependent uniaxial
magneto-crystalline anisotropy (K$_\mathrm{u}$) extracted from ATS
analysis with respective color coding.} \vspace{-1.0em}
\end{figure}

We have fitted the $\emph{m}$(H) isotherms at all the temperatures
with \emph{``ATS model"} for H $\geq$ 50 kOe and determined the
temperature dependence of K$_\mathrm{u}$, as shown in the inset of
Fig. 4. The room temperature K$_\mathrm{u}$ values are 10.24 Merg/cc
and 6.11 Merg/cc for Mn$_3$Ga and Mn$_2$Ga, respectively. The
K$_\mathrm{u}$(T) can be related to H$_\mathrm{c}$ (T) through
modified Stoner-Wohlfarth (SW) relation
\cite{KRONMULLER1988291,GIVORD1990183}, $H_\mathrm{c}(T) = \alpha(T)
\times 2 K_\mathrm{u}(T) / m_\mathrm{sat}(T)$, where $\alpha$(T) is
the microstructural parameter ($\alpha$  = 1 for ideal uniform
magnetization reversal against the anisotropy). The domain wall
movement resulting from the nucleation (with respect to initial
magnetic state) or pinning of magnetic domains in inhomogeneous
regions reduces the anisotropy by the factor $\alpha$(T). The two
mechanism can be distinguished by $\alpha$-parameter, which is $>$
0.3 for nucleation process, whereas $<$ 0.3 for combined pinning and
nucleation processes \cite{KRONMULLER1988291}. Our H$_\mathrm{c}$
(T) analysis using modified SW model gives $\alpha$(T) $\sim$ 0.7
(Mn$_3$Ga) and 0.5 (Mn$_2$Ga) up to T$_\mathrm{c}$. The $\alpha$(T)
values ($>$ 0.3) suggest that the nucleation process is mainly
responsible for hardening of these two alloys. Moreover, the
comparatively lower $\alpha$(T) from Mn$_2$Ga may result from
presence of partial Mn-vacancies (defects) in distorted
ferrimagnetic structure of Mn$_2$Ga. We have also calculated the
characteristic parameter for hard magnet viz. maximum energy
product, $BH_\mathrm{max} = Max(-B \times H)$ in second quadrant.
The room temperature $BH_\mathrm{max}$ values at 80 kOe are 0.61
MGOe and 0.2 MGOe for Mn$_3$Ga and Mn$_2$Ga, respectively. The
thermal variation of $BH_\mathrm{max}$ for two alloys is shown in
inset of Fig. 3, which show a similar behavior to H$_\mathrm{c}$
(T). Here, we point out that the a large discrepancy has been
observed in previously reported room temperature magnetic properties
like m$_\mathrm{s}$, H$_\mathrm{c}$ and $BH_\mathrm{max}$ for
arc-melted alloys
\cite{PhysRevB.77.054406,doi:10.1063/1.362115,QMLu2015,doi:10.1063/1.4759351,PERWEEN2019278,doi:10.1063/1.4942557}.
To understand the differences in our results from previously
reported arc-melted studies, the evolution of chemical and magnetic
structure from Mn$_3$Ga to Mn$_2$Ga has been further investigated
using DFT calculations.

\begin{figure}
\vspace{-1.0em}
\includegraphics[scale=0.27]{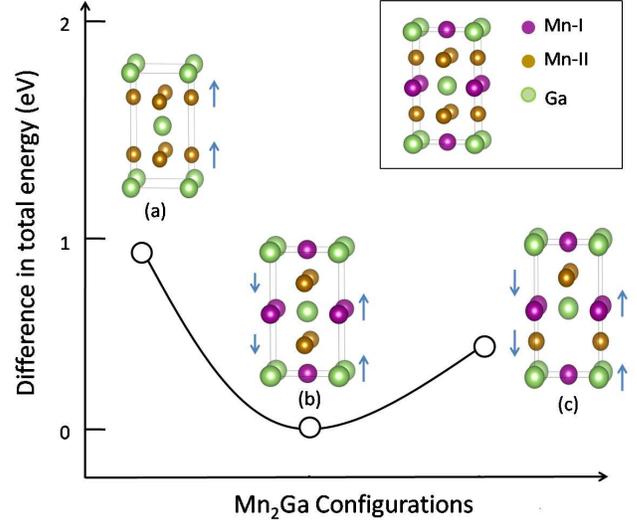}
\vspace{-2.0em} \caption{The calculated total energy difference (per
formula unit) of various Mn$_2$Ga configurations (a,b,c) with
respect to its minimum energy ground state (b), along with the
arrangement of magnetic moments at two (Mn-I and Mn-II) sites. The
unit cell for Mn$_3$Ga is shown in the inset.}
\end{figure}

\begin{figure}
\includegraphics[scale=0.37]{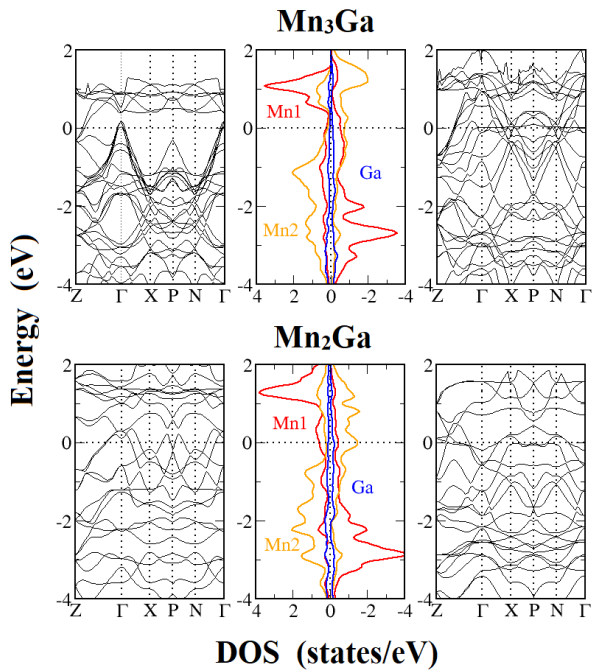}
\vspace{-1.5em} \caption{The atom projected~(color coded) density of
states (middle panel) and the spin up (left panel) and spin down
(right panel) band structure of Mn-Ga alloys. The Fermi level is set
at `zero'.}
\end{figure}

\vspace{-1.0em}
\subsection{Magnetic Ground state: DFT Calculations}
\vspace{-1.0em}

To determine the minimum energy ground state for DO$_{22}$ Mn$_3$Ga
and Mn$_2$Ga alloys, we have carried out DFT calculations by taking
unit cell dimension of experimentally determined lattice parameters.
In D0$_{22}$-type Mn$_3$Ga unit cell with \emph{I4/mmm} symmetry,
Mn-atoms occupy two different Mn-sites, namely, Mn-I at 2b (0, 0,
1/2) and Mn-II at 4d (0, 1/2, 1/4) wyckoff positions with a
multiplicity of 1 and 2, respectively, while Ga atoms only occupy 2a
site [see inset of Fig. 5]. This is in consistent with Heusler alloy
picture \cite{GRAF20111}. In this structure, the magnetic moments at
Mn-I (2b) and Mn-II (4d) sites have antiparallel arrangement,
leading to a ferrimagnetic structure. Next, the Mn$_{3 - \delta}$Ga
($\delta$ $\neq$ 0) unit cell can be realized by removal of Mn from
Mn$_3$Ga unit cell at either (i) Mn-I site, or (ii) Mn-II site or
(iii) both Mn-I and Mn-II sites. Here, we study the minimum energy
ground state of Mn$_2$Ga ($\delta$ = 1) by unraveling several
structural configurations. In Fig. 5, the total energy difference of
three main partially filled DO$_{22}$-type Mn$_2$Ga configurations
with respect to lowest energy configuration is plotted. The complete
removal of Mn-I (2b) atoms [cf. configuration (a)] results in a high
energy configuration with ferromagnetic alignment of magnetic
moments at Mn-II (4d) sites (\emph{P4/mmm} symmetry). In comparison,
the removal of Mn-atoms only from Mn-II (4d) sites results in lower
energy [cf. configurations (b) and (c)], leading to a \emph{I4/mmm}
symmetry for Mn$_2$Ga also; this is in consistent with half-Heusler
alloy structure \cite{GRAF20111}. Out of several such possibilities,
the configuration (b) in Fig. 5 represents the minimum energy
(ground) state for D0$_{22}$ Mn$_2$Ga with \emph{I4/mmm} symmetry.

The magnetic moments associated with DO$_{22}$-type Mn$_3$Ga and
Mn$_2$Ga ground states have been extracted from the total and atom
projected density of states (DOS) calculations [see Fig. 6]. In all
cases, the Mn atoms mainly contribute to the magnetic moment in
these compounds, as evident from the prominent split in the spin up
($\uparrow$) and spin down ($\downarrow$) contribution of the Mn
atoms in the DOS near the Fermi level. The symmetrically
inequivalent Mn atoms at 2b and 4d sites lead to different
(opposite) magnetic contributions. In Mn$_3$Ga configuration, the
net atomic magnetic moments (m$_s$) of 2.841 $\mu_B$ at Mn-I (2b)
and 2.307 $\mu_\mathrm{B}$ at Mn-II (4d) sites lead to an effective
magnetic moment, m = 2 m$_{II}$ - m$_{I}$ - m$_{Ga}$ = 1.74
$\mu$$_B$/f.u. for the system. Whereas, with the removal of a Mn-II
(4d) atoms in Mn$_2$Ga unit cell, the (effective) major contribution
comes from Mn-I (2b) site (3.207 $\mu$$_B$) in comparison to Mn-II
(4d) site (1.837 $\mu$$_B$), resulting in a net magnetic moment, m =
m$_{I}$ - m$_{II}$ + m$_{Ga}$ = 1.39 $\mu$$_B$/f.u. for Mn$_2$Ga.
The inclusion of spin-orbit coupling (SOC) leads to a small
increment in net magnetic moment for two alloys. Furthermore, the
spin polarization for two alloys is calculated as 58\% (53\%) for
Mn$_3$Ga (Mn$_2$Ga), which is in agreement with the experimental
study \cite{doi:10.1002/pssb.201147122}. The summarized atom
specific and net magnetic moments in Mn$_3$Ga and Mn$_2$Ga alloys
without (W/O) and with (W) SOC along with spin polarization are
enlisted in Table I. Thus, the DFT calculations clearly shows that
Mn$_2$Ga adopts a half Heusler structure with removal of only Mn-II
(4d) atoms from full Heusler Mn$_3$Ga unit cell, leading to a
significant decrease in net magnetic moment in comparison to
Mn$_3$Ga. This supports our experimental results on SPS DO$_{22}$
Mn$_{3 - \delta}$Ga ($\delta$ = 0, 1) alloys.

\begin{table}
\begin{ruledtabular}
\begin{tabular}{c|ccc|cc}
System           & \multicolumn {3}{c} {Atomic Moments in $\mu_B$}     &   \multicolumn {2}{c} {Total moment in}\\
                & \multicolumn {3}{c} {(Multiplicity)}                  &   \multicolumn {2}{c} {$\mu_B$/f.u.} \\
\hline
                &    Mn-I   &    Mn-II    &   Ga        &  W/O SOC  &  W SOC        \\
\hline
Mn$_3$Ga        & -2.841(1)  &   2.307(2)  &  -0.065(1)   &   1.74 & 1.8  \\
Mn$_2$Ga        & -3.207(1) &    1.837(2)  &  -0.010(1)    &  1.39 & 1.46 \\
\end{tabular}
\end{ruledtabular}
\caption{The atom specific and net magnetic moments per f.u. with
(W) and without (W/O) spin orbit coupling (SOC), and spin
polarization of Mn$_3$Ga and Mn$_2$Ga alloys.} \vspace{-1.5em}
\end{table}

\vspace{-1.0em}
\section{Conclusion}
\vspace{-1.0em}

In summary, we synthesized the Mn$_{3 - \delta}$Ga ($\delta$ = 0, 1)
alloys with D0$_{22}$ tetragonal structure by Spark Plasma sintering
technique. The magnetization measurements reveal T$_\mathrm{c}$
$\simeq$ 780 K and $\simeq$ 710 K for Mn$_3$Ga and Mn$_2$Ga,
respectively. The magnetic hysteresis loops for Mn$_3$Ga (Mn$_2$Ga)
yield large room temperature spontaneous magnetization
m$_\mathrm{s}$ of 1.63 (0.83) $\mu_\mathrm{B}$/f.u. at 80 kOe,
BH$_\mathrm{max}$ of 0.61 (0.20) MGOe as well as coercivity,
H$_\mathrm{c}$ of 4.285 (3.35) kOe. The Stoner-Wohlfarth model
reveals that the magnetic reversal in these alloys are dominated by
nucleation mechanism. Furthermore, we employ DFT calculations to
identify the ground state structures and the associated magnetic
properties of D0$_{22}$  Mn$_{3 - \delta}$Ga ($\delta$ = 0, 1)
alloys, which are in support of our experimental results. Finally,
we point out that the SPS technique offers a controlled synthesis of
D0$_{22}$ Mn$_{3 - \delta}$Ga (0 $\ge$ $\delta$ $\ge$ 1) alloys,
preserving the Heusler alloy structure.

\bibliography{MnGaSPS-arxiv-ref}

\end{document}